\documentstyle[12pt]{article}
\font\bfit=cmbx10 scaled \magstep1
\def\refmark#1{$^{\themyref}$\addtocounter{myref}{1}}
\def\refshow#1#2{#1$^{\themyref}$\addtocounter{myref}{1}}
\def\refrep#1#2{$^{#2}$}
\def\refrepshow#1#2{#1$^{#2}$}
\def\msol{M_{\odot}}
\def\mdot{\dot M}
\def\lsim{\lower .5ex\hbox{$\buildrel < \over {\sim}$}}
\def\etal{{\it et al.}\ }

\newcommand{\papertitle}[1]{\null\vskip26pt\centerline{\LARGE\bf #1}}
\newcommand{\papertitletwo}[1]{\centerline{\LARGE\bf #1}}
\newcommand{\paperauthor}[1]{\vskip28pt\centerline{\uppercase{#1}}}

\newcommand{\affil}[3]{\vskip 26pt\centerline{\it #1}
     \centerline{\it #2}\centerline{\it #3}}

\newcommand{\heading}[1]{\vskip13pt\centerline{\bf\uppercase{#1}}\vskip13pt}
\newcommand{\headingtwo}[1]{\vskip-13pt\centerline{\bf\uppercase{#1}}\vskip13pt}
\newcommand{\subheading}[1]{\vskip6pt\centerline{\bfit #1}\vskip6pt}
\newcounter{myref}
\begin{document}

\papertitle{The Physics of Core-Collapse}
\smallskip
\papertitletwo{Supernova Explosions\footnote{The authors would like to
acknowledge the NSF both for support under grant \# AST92-17322 and for the
use of NSF Supercomputer Centers where much of the heavy lifting was
performed.}}
\paperauthor{Adam Burrows and John Hayes}
\affil{Departments of Physics and Astronomy}{University of Arizona}{Tucson,
Arizona 85721}
\vskip28pt

\heading{Introduction}
Fresh insights and powerful numerical tools are revitalizing the theoretical
exploration of the supernova mechanism. The realization that the protoneutron
star is Rayleigh-Taylor unstable at various times and radii and, hence, that a
multi-dimensional perspective is required is one agent of this
revolution. However, a new physical understanding of the nature of explosions
(even spherical explosions) that are driven by neutrino heating and that
escape from deep within a gravitational potential well is also
emerging.\refmark{Burrows and Goshy 1993 (BG)}$^,$\refmark{Burrows, Hayes, and
Fryxell 1994 (BHF)}$^,$\refmark{Janka and M\"uller
1993}$^,$\refmark{1994} This,  together with
the new multi-dimensional approach, promises to establish a new paradigm
within which supernova explosions and their consequences can be studied in the
future.

Supernova theory is in flux and a consistent model that fits the growing list
of observational constraints does not yet exist. Nevertheless, the observed
explosion energies, nickel yields,
optical and IR line profiles, pulsar kicks, neutron star masses, nickel debris
distributions, and nucleosynthesis are strengthening the connections between
collapse theory and empirical astronomy. Two of the remaining embarrasments of
theory concern the overproduction of neutron-rich species and the difficulty
of achieving entropies sufficient to produce an r-process. However, recent
simulations hint at how these problems can be solved.\refrep{(BHF)}{2} In
addition, the observation of high-speed pulsars suggests that asymmetries
during and/or after core collapse might exist. Recent calculations by
\refshow{Bazan and Arnett}{(1994)} show that silicon- and oxygen-burning are
hydrodynamic and
that Mach-number and density variations in the core at collapse can be
high. These asymmetries can amplify during infall and can result in asymmetric
explosions that blow preferentially via the paths of least
resistance. \refshow{Burrows and Hayes}{(1995)} have recently demonstrated, via
a 180$^{\circ}$ 2-D hydrodynamic simulation, such a ``rocket'' effect. The
asymmetrically ejected matter causes the residual core to recoil with speeds
of hundreds of kilometers per second. (This suggests that magnetic field
effects are not necessary to achieve high neutron star speeds.) In addition,
there may be a correlation between the pulsar recoils and the distribution of
the ejected $^{56}$Ni.

There are far too many new and interesting questions, problems, and potential
solutions to be comfortably contained in this short paper. Therefore, we will
focus here on a discussion of the crucial ingredients of the explosion {\it
mechanism} itself and the character of the blast after it starts. A more
in-depth discussion of some of the constraints listed in Table 1 and the 1-D
and 2-D hydrodynamic simulations that we have recently performed can be found
in \refrepshow{BHF}{2}. Color figures from that paper and mpeg movies of one
of its
2-D simulations can be acquired gratis via mosaic at URL address
http://lepton.physics.arizona.edu:8000/.

\heading{Neutrino-Driven Explosions}
\headingtwo{in One and Two Dimensions}

\subheading{a. The Quasi-Steady-State Phase Before Explosion and the Explosion
Condition}

After the ``Chandrasekhar'' core of a massive star becomes unstable to
implosion, it evolves through various distinct hydrodynamic phases. These are
infall, core bounce, shock formation, shock stagnation, the pre-explosion
quasi-steady state, the onset of explosion, and the explosion proper. Twenty
to one hundred milliseconds into the explosion, a distinct neutrino-driven
wind emerges from the core, whatever the details of the mechanism. Since the
direct hydrodynamic mechanism aborts for all progenitors (even the lightest
massive stars), the nature and evolution of the quasi-steady state after the
shock stalls takes on a new importance. How long does the steady-state phase
last? What triggers the explosion? How does the explosion evolve? In what
context does a black hole form? The answers to all these questions hinge on
the proper understanding of the physics of the shock-bounded and accreting
protoneutron star.

\refrepshow{Burrows and Goshy (1993, BG)}{1} have recently developed an
approximate semi-analytic theory of such objects. Setting all partial
derivatives with
respect to time equal to zero, the equations of hydrodynamics and neutrino
transfer become a set of coupled ordinary differential equations, subject to
boundary conditions. Such a problem is an {\it eigenvalue} problem. With an
equation of state, a prescription for the bounding shock jump conditions and
the outer supersonic flow profiles, a given core mass $(\sim1.1$--$1.3\msol)$,
and simple formulae for neutrino heating and cooling exterior to the
neutrinospheres, \refrepshow{BG}{1} solved for the steady-state structure. The
shock
radius $(R_s)$ was the eigenvalue and the electron neutrino luminosity
$(L_{\nu_e})$ and the mass accretion rate $(\mdot)$ were the control
parameters. Physically, the structure adjusts until the infall time from the
shock to the core $({R_s\over u_1}$, where $u_1$ is the post-shock settling
velocity) is ``equal'' to the cooling or heating timescale. This equality of
timescales is similar to the equality of the free-fall and sound travel times
in the context of hydrostatic equilibrium and to a similar condition in the
context of AM Her stand-off shocks.\refmark{Chevalier and Imamura
1982}$^,$\refmark{Langer \etal 1981} The quasi-steady assumption is proper as
long as
these adjustment timescales ($\sim$10 milliseconds) are shorter than the
timescale for the decay of $\mdot$ (30-100 milliseconds). Note that the
radius to which the bounce shock is initially thrown is almost unrelated to
its later steady values.

Increasing $L_{\nu_e}$ for a given $\mdot$, increases $R_s$ (roughly as
$L^2_{\nu_e}/\mdot$). However, this behavior does not continue for arbitrarily
high $L_{\nu_e}$. \refrepshow{BG}{1} showed that for each $\mdot$ (and set of
model
assumptions), there is a {\it critical} $L_{\nu_e}$ above which there is no
steady-state protoneutron star envelope (Figure 1). This is reached at a
finite value of the eignevalue, $R_s$ (generally less than 200
kilometers). \refrepshow{BG}{1} identified the implied {\it instability} with
the
onset of the supernova explosion. This onset is a critical phenomenon and is
at the bifurcation between steady-state and wind solutions of the
equations. If,
for a given $\mdot$, $L_{\nu_e}$ could be increased by better neutrino
transport or cross sections or by convective enhancement,\refmark{Burrows
1987}$^,$\refmark{Mayle and Wilson 1988} the 1-D models would explode more
readily. The problem with previous 1-D calculations is that their
$L_{\nu_e}$--$\mdot$ trajectories passed below the critical curve, as Figure 1
depicts.

Recently, it was shown that the outer shocked envelopes of the protoneutron
star are generically unstable to Rayleigh-Taylor overturn driven by neutrino
heating from below.\refmark{Herant, Benz and Colgate 1992}$^,$\refmark{Herant
\etal 1994 (HBHFC)}$^,$\refmark{Bethe
1990}$^,$\refrep{BHF}{2}$^,$\refmark{Janka and
M\"uller, this volume} This and other hydrodynamic instabilities before,
during, and after explosion are redrawing our picture of the evolution and
character of supernova blasts. An important question one may ask is: how do the
multi-dimensional effects alter the explosion mechanism? On this there is much
needless confusion that the next paragraphs may partially clear up.

Figure 2 depicts in cartoon form the shock-bounded protoneutron star before
explosion in 1-D and $\ge$2-D. Relaxing spherical symmetry allows some parcels
of matter that have just passed through the shock and that are being heated by
neutrinos from the core to rise like balloons or cells in any ``convectively''
unstable region. This allows the matter to dwell longer in the gain region
(where heating $>$ cooling) and, hence, to achieve higher entropies than is
possible in 1-D.\refrep{HBC}{11}$^,$\refrep{Bethe 1990}{13}

\vbox{\vskip21pc
{\small
\noindent\uppercase{Figure 1}.
Approximate critical curves for explosion, with and without multi-dimensional
effects, in $L_{\nu_e}$ versus $\mdot$ space. Superposed is a representative
$L_{\nu_e}$ vs.\ $\mdot$ trajectory for a realistic calculation. Note that
this curve intersects the lower $\ge 2$-D critical curve at some time, whereas
in 1-D it may never.}}
\newpage

\vbox{\vskip25truepc
{\small
\noindent \uppercase{Figure 2}.
Schematics of the accreting protoneutron star in 1- and
$\ge$2-dimensions. The gain region is indicated by the hatches. The main
difference between the two is the larger steady-state radius that overturn
allows.}}
\bigskip

In 1-D, the heated
parcel would perforce fall directly into the cooling region interior to
the gain radius (where cooling = heating) and lose its just recently
gained energy. (Curiously, the cooling region lies closer to the
neutrinospheres where the temperatures are higher.)

However, the rising balloons, upon
encountering the shock, are immediately advected inward by the powerful mass
accretion flux raining down. They do not dwell near the shock. In fact, the
net mass flux through the shock is approximately equal to the mass flux onto
the core and mass does not accumulate in the convective zone. The mass between
the shock and the neutrinospheres {\it decreased} by about a factor of three
in the calculations of \refrepshow{BHF}{2} during the pre-explosion boiling
phase
that lasted $\sim$100 milliseconds ($\sim$30 convective turnover times). {\it
All} the matter that participates in the ``convection'' before explosion
eventually leaves the convection zone and settles onto the core. A given
parcel of matter may ``cycle'' one or two times before settling inward (and a
large fraction never rises), but more than three times is rare. The boiling
zone is resupplied with mass by mass accretion through the shock and a
secularly evolving steady-state is reached. This steady-state is similar to
that achieved in 1-D, but due to the higher dwell time the average entropy in
the envelope is larger and its entropy gradient is flatter. These effects,
together with the dynamical pressure of the buoyant plumes, serve to increase
the steady-state shock radius over its value in 1-D by 30\%--100\%. It is this
effect of boiling that is central to its role in triggering the explosion, for
it thereby lowers the critical luminosity threshold. As Figure 1 suggests, the
lowering of the effective critical curve allows the actual model trajectory in
$L_{\nu_e}$ vs. $\mdot$ space to intersect it. Even if in 1-D it can be shown
that the two curves can intersect, they would intersect earlier and more
assuredly with the multi-dimensional effects included. The physical reasons
for the lowered threshold are straightforward: a large $R_s$ enlarges the
volume of the gain region, puts shocked matter lower in the gravitational
potential well, and lowers the accretion ram pressure at the shock for a given
$\mdot$. Since the ``escape'' temperature $(T_{\rm esc}\propto{GM\mu\over kR})$
decreases with radius faster than the actual matter temperature $(T)$ behind
the shock, a larger $R_s$ puts a larger fraction of the shocked material above
its local escape temperature. $T>T_{\rm esc}$ is the condition for
a thermally-driven corona to lift off of a star. In one, two, or three
dimensions, since supernovae are driven by neutrino heating, they are coronal
phenomena, akin to winds, though initially bounded by an accretion
tamp. Neutrino radiation pressure is unimportant.

We conclude that the instability that leads to explosion in $\ge$2-D is of the
same character as that which leads to explosion in 1-D. Since the explosion
succeeds a quasi-steady-state phase, neither the total neutrino energy
deposited during the boiling phase nor any putative coeval thermodynamic cycle
is of relevance to the energy of the explosion or the trigger
criterion. Energy does not accumulate in the overturning region before
explosion (it in fact decreases) and the increasing vigor (speed) of
convection is in response to the
decay of $\mdot$. If $\mdot$ were held constant, the overturning would not
grow more vigorous with every ``cycle'' and a simple, stable convective zone
would be established. In fact, before explosion the average total energy
fluxes $((\epsilon+P/\rho+{1\over 2}v^2-{GM\over r})\mdot)$ due to the
overturning motions are {\it inward}, not outward, since the net direction of
the matter is onto the core. Figure 3 depicts such fluxes versus radius at
various times for the 1-D and 2-D simulations conducted by \refrepshow{BHF}{2}
of
the core of a 15$\msol$ star. The hump on the inside mirrors the corresponding
plot for the total neutrino luminosities. During the boiling phase, the net
fluxes in the gain region are negative, not positive, and they become positive
only after the explosion commences. This interpretation of the role of 2-D and
the nature of supernova explosions differs from that of \refrepshow{HBC}{11}
and \refrepshow{HBHFC}{12}.

\subheading{b. The Explosion}
Importantly, just after the explosion criterion is achieved, the explosion
energy is still not determined. In fact, the matter that will eventually be
ejected is often still {\it bound}, even correcting for the reassociation
boost (the ``after-burner''). This fact emphasizes that the explosion
condition has nothing to do with the aggregate neutrino deposition before
explosion and that this aggregate heating has nothing to do with the explosion
energy itself. The onset of explosion causes the shocked matter to expand and
to lower its temperature, thereby turning off cooling. The expansion runs away
because heating continues unchecked by cooling, whose integral had usually
dominated between the neutrinosphere and the shock during the quasi-state
phase. The explosion and the shock are thereafter {\it driven} by neutrino
energy deposition. This continuing source is necessary so that the ejecta can
eventually achieve positive energies of supernova magnitude. Soon after the
explosion commences, all convective cycling between the shock and
neutrinospheres ceases permanently. Note that it is only after the shock has
achieved many thousands of kilometers that the explosion has had a chance to
``feel'' the mantle binding energy that it will eventually have to overcome to
succeed. If this binding energy is large, the explosion will be slow. This
will allow the material of the explosion to be heated longer. This feedback
effect partially compensates for the variety of envelope binding energies
along the
massive star continuum,\refrep{BHF}{2} and may explain why supernova energies
are all near $10^{51}$ ergs. Binding energies of massive star envelopes are of
order $10^{51}$ ergs because this is approximately the binding energy of a
Chandrasekhar white dwarf $(\sim m_ec^2{\rm N}_{\rm A}{\rm M}_{\rm
CH})$. (Simple arguments show that the envelope and the core binding energies
are comparable). Therefore, and very crudely, the envelope binding energies
set the scale of the explosion energy (to within a factor of three?), though
we still can't say whether it increases or decreases with ZAMS mass. It is
suspected that the envelope binding energies can be too high and that the
accretion tamp can be too oppressive and that above some ZAMS mass, the core
will collapse to a black hole before or during explosion. It is not known
whether, during the first seconds after bounce, a supernova and a black hole
are
mutually exclusive results.
(We know from its neutrino signal that in SN1987A the neutron star lasted at
least 12.5 seconds.)

\medskip
\vbox{\vskip29truepc
{\small
\noindent \uppercase{Figure 3}.
The angle-averaged total energy fluxes versus radius in the one- and
two-dimensional calculations of BHF. The average position of the shock is
identified with a circle. Note that the energy fluxes are not outward in 2-D,
until the explosion begins.}}

\heading{Discussion}
Though we have made progress in understanding supernova explosions, we still
don't know the interesting systematics as a function of progenitor mass. What
are $E_{SN}(M_{\rm ZAMS})$, $M_{NS}(M_{\rm ZAMS})$, the critical mass for
black hole formation, and the r-process and $^{56}$Ni yields? In addition, all
the 2-D explosion simulations to date eject too much neutron-rich matter. This
problem may be solved if the delay to explosion ($\sim$50 milliseconds for
\refrepshow{HBHFC}{12} and $\sim$100 milliseconds for \refrepshow{BHF}{2})
were longer still, allowing the
envelope mass to decrease and the density of the accreting matter to
thin. (The latter is important because electron-capture rates are stiffly
increasing functions of density.) Thus far, no multi-dimensional explosion
simulation has involved multi-group transport, general relativity, or been
done in 3-D. We are only at the beginning of a new theoretical assault on the
supernova phenomenon that will provide scores of research projects for this
and the next generation of modelers.

\heading{References}
\let\uc=\uppercase

\begin{enumerate}

\item \uc{Burrows, A., J. Goshy.} 1993. Ap.~J.~Lett. {\bf 416}:L75 (BG).

\item \uc{Burrows, A., J. Hayes, and B. A. Fryxell}. 1994. submitted to
Ap. J. (BHF).

\item \uc{Janka, H.-T., E. M\"uller}. 1993. In the proceedings of the
Internat'l Symposium on Neutrino Astrophysics, held in Takayama/Kamioka Japan
Oct. 19--22, 1992, to appear in {\it Frontiers of Neutrino Astrophysics},
(Universal Academy Press Inc. Tokyo, Japan).

\item \uc{Janka, H. T., E. M\"uller}. 1994. A\&A in press.

\item \uc{Bazan, G., and D. Arnett}. 1994. Ap.~J.~Lett. {\bf 433}:L41.

\item \uc{Burrows, A., and J. Hayes}. 1995. in preparation.

\item \uc{Chevalier, R. A., and J. N. Imamura}. 1982. Ap.~J. {\bf 261}:543.

\item \uc{Langer, S. H., G. Chanmugam, and G. Shaviv}. 1981. Ap.~J. {\bf
245}:L23.

\item \uc{Burrows, A}. 1987. Ap.~J.~Lett. {\bf 318}:L57.

\item \uc{Mayle, R., J. R. Wilson}. 1988. Ap.~J. {\bf 334}:909.

\item \uc{Herant, M., W. Benz. S. A. Colgate}. 1992. Ap.~J. {\bf 395}:642
(HBC).

\item \uc{Herant, M., W. Benz, J. Hix, C. Fryer, and
S. A. Colgate}. 1994. Ap.~J. {\bf 435}: 339 (HBHFC).

\item \uc{Bethe, H}. 1990. Rev.~Mod.~Phys. {\bf 62}:801.

\item \uc{Janka, H. T., E. M\"uller}. This volume.

\end{enumerate}

\end{document}